\documentclass[aps,pre,floats,twocolumn,nofootinbib,showpacs]{revtex4}

\usepackage{epsfig}
\usepackage{graphicx}
\usepackage{latexsym,amsmath,verbatim}

\newcommand{\avn}[1]{\langle k^{#1}\rangle}
\newcommand{\fluck}{\langle k^2\rangle}
\newcommand{\avk}{\langle k\rangle}

\begin{document}

\title{Bosonic reaction-diffusion processes
on scale-free networks}

\author{Andrea Baronchelli, Michele Catanzaro and Romualdo
  Pastor-Satorras} 
\affiliation{Departament de F\'\i sica i Enginyeria Nuclear,
  Universitat Polit\`ecnica de Catalunya, Campus Nord B4, 08034
  Barcelona, Spain}
\date{\today}

\begin{abstract}
  Reaction-diffusion processes can be adopted 
  to model a large number of dynamics on complex networks, 
  such as transport processes or epidemic outbreaks. 
  In most cases, however, they have been studied from a
  fermionic perspective, in which each vertex can be occupied by at
  most one particle.  While still useful, this approach suffers
  from some drawbacks, the most important probably being the
  difficulty to implement reactions involving more than two particles
  simultaneously. Here we introduce a general framework for the
  study of bosonic reaction-diffusion processes on complex
  networks, in which there is no restriction on the number of
  interacting particles that a vertex can host.  We describe these
  processes theoretically by means of continuous time heterogeneous
  mean-field theory and divide them into two main classes: steady
  state and monotonously decaying processes. We analyze specific
  examples of both behaviors within the class of one-species process,
  comparing the results (whenever possible) with the corresponding
  fermionic counterparts.  We find that the time evolution and
  critical properties of the particle density are independent of the
  fermionic or bosonic nature of the process, while differences exist
  in the functional form of the density of occupied vertices in a
  given degree class $k$. We implement a continuous time Monte Carlo
  algorithm, well suited for general bosonic simulations, which allow
  us to confirm the analytical predictions formulated within mean-field theory. 
  Our results, both at the theoretical and
  numerical level, can be easily generalized to tackle more complex,
  multi-species, reaction-diffusion processes, and open a
  promising path for a general study and classification of this kind
  of dynamical systems on complex networks.
\end{abstract}


\maketitle

\section{Introduction}

Many natural, social, and artificial systems exhibit heterogeneous
patterns of connections and interactions that can be naturally
described in terms of networks or graphs~\cite{bollobas98}.  Thus,
complex network theory turns out to be the natural framework in which
the functional and structural properties of complex systems belonging
to completely different domains can be rationalized and
investigated~\cite{barabasi02,mendesbook,newman-review,caldarelli_book}. This approach
has recently proved to be very powerful, and systematic statistical
analysis have allowed to recognize the existence of many
characteristic features shared by a large class of different systems,
the most peculiar being the small-world property~\cite{watts98} and a
large connectivity heterogeneity yielding a scale-free degree
distribution~\cite{barab99}.  A graph is said to be small-world when
the average topological distance between any pair of vertices is
``small'', scaling logarithmically or slower with the system size $N$.
On the other hand, defining the degree $k$ of a vertex as the number of
connections linking it to other vertices, scale-free (SF) networks are
characterized by a degree distribution $P(k)$ that decreases as a
power-law,
\begin{equation}
  P(k) \sim k^{-\gamma},
\end{equation}
where $\gamma$ is a characteristic degree exponent, usually in the
range $2 < \gamma < 3$. 

The distinctive structural properties of networked systems, beyond
being intrinsically interesting, have also a strong impact on the
dynamical processes taking place on such
systems~\cite{dorogovtsev07:_critic_phenom}, which can have practical
implications in, e.g., understanding traffic behavior in technological
systems such as the Internet \cite{romuvespibook}.  In particular, the
heterogeneous connectivity pattern of SF networks with
diverging second moment $\fluck$ (i.e., with $\gamma \leq3$) can lead
to very surprising dynamical properties, such as an extreme weakness
in front of targeted attacks, aimed at destroying the most connected
vertices \cite{havlin01,newman00}, as well as to the propagation of
infective agents \cite{pv01a,lloyd01}. After those initial
discoveries, a large series of new results has been put forward and we
refer the reader to
Refs.~\cite{bocaletti06:_compl_networ,dorogovtsev07:_critic_phenom}
for recent reviews on the subject.

A powerful framework to describe many dynamical processes in a most
general way is given by the theory of reaction-diffusion (RD)
processes~\cite{vankampen}. RD processes are defined in terms of
different kinds of particles or ``species'', which diffuse
stochastically (usually by performing a random walk) and interact
among them according to a given set of reaction rules. Apart from
their natural application to describe chemical reactions, RD processes
are useful to represent any system in which different kinds of
``agents'' diffuse in space and dissappear, are created, or change
their state, according to the state of other agents in a given
neighborhood.  An example of this kind of processes is the spread of
diseases in population systems.  An epidemic process leading to an
endemic state can be described by the Susceptible-Infected-Susceptible
(SIS) model~\cite{epidemics}, which corresponds to an RD process with
two species (individuals), representing susceptible ($S$) and infected
($I$) individuals, which diffuse (with possible different diffusion
rates) and interact through the reactions
\begin{equation}
  \begin{array}{ccc}
  S + I &\to& 2 I, \\
  I &\to& S,
  \end{array}
  \label{eq:25}
\end{equation}
representing susceptible individuals becoming infected upon
encountering an infected individual, and infected individuals
spontaneously recovering.

RD processes on regular topologies (Euclidean lattices) have been
extensively studied, and an elegant formalism has been developed to
allow for a general description in terms of field
theories~\cite{RevModPhys.70.979}. On the other hand, the effects of
more complex, heterogeneous, topologies have been taken into account
only recently for simple
processes~\cite{originalA+A,michelediffusion,nohkim,weber:046108,ke06:_migrat_driven,chang07:_react_diffus_proces},
and a systematic description of this interesting problem is still
lacking. Moreover, so far most of the attention has been devoted to
the restricted case of \textit{fermionic} (or \textit{microscopic}, in
the chemistry jargon) RD processes, in which a vertex of the network
cannot be occupied by more than one particle. In this context,
numerical and analytical results have been put forward for the most
simple RD processes, namely the
diffusion-annihilation~\cite{originalA+A,michelediffusion} and the
diffusion-coagulation \cite{originalA+A,weber:046108} processes.
Although these results are undoubtedly interesting and offer an
initial insight into the behavior of RD processes in heterogeneous
networks, the adopted fermionic approach suffers from two considerable
conceptual drawbacks: (i) there is no systematic framework for the
description of this kind of processes, and both numerical models and
theoretical approximations (through heterogeneous mean-field theory)
must be considered on a case by cases basis; and (ii) it is relatively
easy to deal with RD processes with at most order-two reactions
(involving at most two particles), but it becomes more problematic to
implement reactions among three or more particles.  Thus, for example,
a fermionic study of the three particles reaction $A+B+C \to
\emptyset$ \cite{chang07:_react_diffus_proces} requires the
introduction of an artificial ``intermediate'' particle, created from
the reaction of two particles, and that reacts itself with a third,
leading to the actual annihilation event.  In this sense, it seems
more natural and realistic to consider instead \textit{bosonic} (or
\textit{mesoscopic}) processes, in which there are no restrictions on
the vertex occupancy, and for which, levering in what is already known
for Euclidean lattices, it is possible to develop systematic
analytical and numerical formalisms. Moreover, while some processes
may naturally fit into a fermionic framework, other are intrinsically
bosonic. For example, during the spreading of a disease (say HIV) on a
social interaction network, each individual can only change its state
(become infected) by contagion through one acquaintance in the
network. However, the diffusion of a disease at the level of airport
networks (for example, SARS) is better modeled by taking into account
the number of infected individuals in each city
\cite{colizza06:_predic}. The bosonic version of RD processes on
complex networks has been so far neglected, with the exception of
Ref.~\cite{v.07:_react}, where it has been applied to the particular
case of the SIS process, Eq.~(\ref{eq:25}) (see also
Ref.~\cite{colizza07:_invas_thres} for an extension of the model to
weighted networks).

In this paper, we investigate the properties of general bosonic RD
processes in complex heterogeneous networks, adopting a twofold
continuous time approach based on heterogeneous mean-field (MF) theory
and numerical simulations. We develop a general MF formalism, based on
the standard law of mass action, that is able to describe any RD
processes on general complex networks, taking a particularly simple
form in one-species RD systems. For this case, general predictions,
independent of the particular form of the reaction rules, can be made
in the small particle density (diffusion-limited) regime. The
formalism is applied and fully solved in two particular cases, the
branching-annihilating random walk and the diffusion-annihilation
problem, examples of RD systems with stationary states and
monotonously decaying particle densities, respectively. In order to
check the possible differences between the bosonic and fermionic
implementations of the same problem, we consider at the same time both
examples from the fermionic MF theory perspective. We find that both
formalisms provide analogous results for the time evolution and
critical properties of the dynamics. However, the two approaches are
not completely equivalent: the functional form of the particle density
restricted to vertices of given degree $k$ varies widely between the
two approaches.  Finally we check our results, and in particular the
equivalence between fermionic and bosonic formalisms, by means of
extensive computer simulations. Contrarily to previous
approaches~\cite{v.07:_react}, in which a parallel updating scheme was
defined for the particular model under scrutiny, 
we adopt a sequential continuous time algorithm that can be easily
generalized for any RD process.

The paper is organized as follows.  In Sec.~\ref{sec:react-diff-proc}
we define general bosonic RD processes in complex networks.
Sec.~\ref{sec:mean-field-formalism} is devoted to the introduction of
a generic analytical framework based on bosonic heterogeneous MF
formalism, from which general predictions can be obtained for any kind
of RD process.  In Sec.~\ref{sec:applications} we consider and solve
some particular examples of RD processes, exhibiting steady states and
a monotonously decaying density, namely the branching annihilating
random walk and diffusion-annihilation processes, respectively.  The
predictions of heterogeneous MF theory are validated in
Sec.~\ref{sec:numer-simul} by means of numerical simulations. Finally,
in Sec.~\ref{sec:discussion} we summarize and discuss our results.

\section{Bosonic RD processes in complex networks}
\label{sec:react-diff-proc}

We consider RD processes on complex networks,
which are fully defined by the adjacency matrix $a_{ij}$, which takes
the values $a_{ij}=1$ if vertices $i$ and $j$ are connected by an
edge, and zero otherwise. From a statistical point of view, the
network can also be described by its degree distribution $P(k)$ and its
degree correlations, given by the conditional probability $P(k'|k)$
that a vertex of degree $k$ is connected to a vertex of degree $k'$
\cite{serrano07:_correl}. Both descriptions are related through the
formulas
\begin{equation}
  P(k) = \frac{1}{N} \sum_i \delta(k, \sum_j a_{ij}),
\end{equation}
where $\delta(x,y)$ is the Kronecker $\delta$ symbol, and
\cite{michelediffusion}
\begin{equation}
  P(k'|k) = \frac{1}{N k P(k)} \sum_{i \in k} \sum_{j \in k'} a_{ij},
  \label{eq:22}
\end{equation}
$N$ being the size of the network.

RD processes are defined as dynamical systems involving particles of
$S$ different species $A_\alpha$, $\alpha=1,\ldots,S$, that diffuse
stochastically on the vertices of the network and interact among them
upon contact on the same vertex, following a predefined set of $R$
reaction rules. In a bosonic scheme, there is no limitation in the
number of particles that a vertex can hold, therefore the occupation
numbers $n^\alpha_i(t)$, denoting the number of particles of species
$A_\alpha$ in vertex $i$ at time $t$, can take any value between $0$
and $\infty$. We will assume that diffusion in the network is
homogeneous and takes place by means of random jumps between nearest
neighbors vertices. Therefore, an $A_\alpha$ particle with a diffusion
coefficient $D_\alpha$ at vertex $i$ will jump with a probability per
unit time $D_\alpha /k_i $ to a vertex $j$ adjacent to $i$, where
$k_i$ is the degree of the first vertex.

The reaction rules that particles experience upon contact, on the
other hand, can be defined in the most general way by the
corresponding stoichiometric equations \cite{guggenheim67:_therm}
\begin{equation}
  \sum_{\alpha=1}^S q_\alpha^r A_\alpha
  \stackrel{\lambda_{r}}{\longrightarrow} \sum _{\alpha=1}^S
  (q_\alpha^r+ p_\alpha^r ) A_\alpha,  \quad r=1,\ldots, R,
  \label{eq:17}
\end{equation}
where $q_\alpha^r>0$ (we do not consider reactions involving the
spontaneous creation of particles) and $p_\alpha^r \geq
-q_\alpha^r$. The coefficients $q_\alpha^r$ and $p_\alpha^r$ define
the $r$-th reaction process, while $\lambda_{r}$ is the probability
per unit time that the reaction takes place.  Given that the reactions
take place inside the vertices, the only variation between a RD
process in a complex network and a regular lattices lies in the
diffusion step. As we will see, however, this variation alone can
induce important differences between processes in these two
reaction substrates.

\section{Heterogeneous continuous-time Bosonic mean-field formalism}
\label{sec:mean-field-formalism}

A first analytical description of dynamical processes of complex
networks can be obtained by means of heterogeneous MF theory
\cite{dorogovtsev07:_critic_phenom}. MF theory applied to networks is
based in the assumption that all vertices with the same degree share
essentially the same dynamic properties, and can therefore be
consistently grouped into the same degree class. In the case of RD
processes, and in order to allow for the possibility of network
heterogeneity and large degree fluctuations, it becomes necessary to
work with the density spectra $\rho_{\alpha, k}(t)$
\cite{pv01a,pv01b}, representing the partial density of $A_\alpha$
particles in vertices of degree $k$, and that is defined as
\begin{equation}
  \rho_{\alpha, k}(t) = \frac{\bar{n}_{\alpha, k}(t)}{N_k},
\end{equation}
where $\bar{n}_{\alpha, k}(t)$ is the average occupation number of
particles $A_\alpha$ in the class of vertices of degree $k$ and $N_k=N
P(k)$ is the number of vertices of degree $k$ in a network of size
$N$. From the density spectra, the total density of $A_\alpha$
particles is given by
\begin{equation}
  \rho_\alpha(t) = \sum_k P(k)  \rho_{\alpha, k}(t).
  \label{eq:5}
\end{equation}

Heterogeneous MF theory is given in terms of rate equations for the
variation of the partial densities $\rho_{\alpha, k}(t)$, which in
this case are composed by two terms: one dealing with the (linear)
diffusion and another with the reactions, so we can write
\begin{equation}
  \frac{\partial  \rho_{\alpha, k}(t)}{\partial t} = 
  \mathcal{D_\alpha} +   \mathcal{R}_\alpha.
\label{eq:RD_general}
\end{equation}
The diffusion term is easy to obtain by considering the diffusion
dynamics at the vertex level. The total change of $A_\alpha$ particles
at vertex $i$ is due to the outflow of particles jumping out at rate
$D_\alpha$, plus the inflow corresponding to jumps of particles from
nearest neighbors. Therefore, the diffusive component at the single
vertex level satisfies the rate equation \cite{michelediffusion}
\begin{equation}
  \frac{\partial  n_{\alpha, i}(t)}{\partial t} = -D_\alpha n_{\alpha,
    i}(t)
  + D_\alpha \sum_j 
  \frac{a_{ij}}{k_j} n_{\alpha, j}(t).
\end{equation}
Considering the density spectrum  as the average
\begin{equation}
  \rho_{\alpha, k}(t) = \frac{\sum_{i \in k} n_{\alpha, i}}{N_k}
\end{equation}
and assuming that $n_{\alpha, i}(t) \simeq \bar{n}_{\alpha, k}(t)$, $\forall
i \in k$, we obtain
\begin{equation}
  \mathcal{D}_\alpha  = -D_\alpha \rho_{\alpha, k}(t) +
  D_\alpha k \sum_{k'} 
  \frac{P(k'|k)}{k'} \rho_{\alpha, k'}(t),
\end{equation}
where we have used Eq.~(\ref{eq:22}).

The reaction term can be directly derived from the law of mass action,
according to which the rate of any (chemical) reaction is proportional
to the product of the concentrations (or densities) of the
reactants~\cite{gardiner}.  Considering the set of all allowed
processes Eq.~(\ref{eq:17}), we obtain:
\begin{equation}
  \mathcal{R}_\alpha = \sum_{r} p_\alpha^r  \lambda_{r} 
  \prod_\beta [\rho_{\beta, k}(t)]^{q_\beta^r}.
\end{equation}
Collecting all terms, the rate equations for the density spectra can
be written in the most general case as
\begin{eqnarray}
  \frac{\partial  \rho_{\alpha, k} (t)}{\partial t} &=& -D_\alpha
\rho_{\alpha,k}(t) +
  D_\alpha k \sum_{k'} 
  \frac{P(k'|k)}{k'} \rho_{\alpha, k'}(t) \nonumber \\
  &+& 
\sum_{r} p_\alpha^r  \lambda_{r} 
  \prod_\beta [\rho_{\beta, k}(t)]^{q_\beta^r},
  \label{eq:18}
\end{eqnarray}
while the total densities satisfy the equations
\begin{equation}
  \frac{\partial  \rho_\alpha (t)}{\partial t} =  
  \sum_{r} p_\alpha^r \lambda_{r} 
  \sum_k P(k)\prod_\beta [\rho_{\beta,k}(t)]^{q_\beta^r},
  \label{eq:19}
\end{equation}
where we have used the degree detailed balance condition
\cite{marian1}
\begin{equation}
  kP(k)P(k'|k)=k'P(k')P(k|k').
  \label{detailed_balance}
\end{equation}
It is noteworthy that Eq.~(\ref{eq:19}) is explicitly independent of
the particular form of the network's degree correlations, which only
appear implicitly through the form of the density spectra
$\rho_{\alpha,k}$.

In the following, we will focus in the analysis of one-species RD
processes, in which a single class of particles diffuse and react in
the system, i.e. $S=1$. In this
case, reactions of the same order can be grouped, and
Eqs.~(\ref{eq:18}) and~(\ref{eq:19}) take the simpler forms, omitting
the $\alpha$ index, 
\begin{eqnarray}
  \frac{\partial  \rho_k (t)}{\partial t} &=& 
  k \sum_{k'} 
  \frac{P(k'|k)}{k'} \rho_{k'}(t) + \sum_{q>0} \Gamma_q
  [\rho_k(t)]^q, \label{eq:20}\\
   \frac{\partial  \rho(t)}{\partial t} &=&  \rho(t) + \sum_{q>0}
   \Gamma_q \sum_k
   P(k)  [\rho_k(t)]^q,\label{eq:21} 
\end{eqnarray}
where
\begin{equation}
  \Gamma_q =- \delta(q,1) +  \sum_r p^r \lambda_r \delta(q^r, q),
\end{equation}
and we have absorbed the diffusion rate $D$ into a redefinition of the
time scale and the reaction rates $\lambda_r$.

RD processes with non diverging solutions for Eqs.~(\ref{eq:20})
and~(\ref{eq:21}) can be generally grouped in two classes: those
yielding a particle density monotonously decaying in time and those
exhibiting one or more steady states, with possibly associated phase
transitions between different steady states. We will examine more
closely these two cases in the following subsections. While a full
theoretical analysis requires detailed information about the
particular form of the reactions involved and the network's
degree correlations, it is possible, however, to make very general
statements, and to obtain the asymptotic form of the solutions when
the particle density $\rho$ is very small.

\subsection{Steady-state  Bosonic RD processes}
\label{sec:steady-state-rd-1}

RD processes with steady states possess nonzero solutions for the long
time limit of Eq.~(\ref{eq:20}). In particular, imposing
$\partial_t\rho_k =0$, the steady states correspond to the solutions
of the algebraic equation
\begin{equation}
  \rho_k = - \frac{k}{\Gamma_1} \sum_{k'} \frac{P(k'|k)}{k'}
  \rho_{k'}(t) - \sum_{q>1} \frac{\Gamma_q}{\Gamma_1}  [\rho_k]^q, 
    \label{eq:4}
\end{equation}
where we assume $\Gamma_1 \neq 0$. Since we do not consider the
spontaneous creation of particles from void ($\Gamma_0=0$), $\rho_k=0$
is a solution of Eq.~(\ref{eq:4}). This equation is extremely
difficult to solve for a general correlation pattern $P(k'|k)$, in
order to find nonzero solutions. The condition for this nonzero
solution to exist, however, can be obtained for any correlation
pattern by performing a linear stability analysis \cite{marian1} in
Eq.~(\ref{eq:20}). Neglecting higher order terms, Eq.~(\ref{eq:20})
becomes
\begin{equation}
  \frac{\partial  \rho_k(t)}{\partial t} \simeq 
  \sum_{k'} L_{k k'}  \rho_{k'}(t), 
\end{equation}
where we have defined the Jacobian matrix
\begin{equation}
  L_{k k'} = \Gamma_1  \delta(k', k) +  \frac{k P(k'|k)}{k'}.
\end{equation}
It is easy to see that this matrix has a unique eigenvector $v_k = k$
and a unique eigenvalue $\Lambda = \Gamma_1+1$. Therefore, defining
$\tilde{\Gamma}_1 = \Gamma_1+1 \equiv \sum_r p^r \lambda_r \delta(q^r,
1)$, a nonzero steady state is only possible when
$\tilde{\Gamma}_1>0$, which translates in the presence of reaction
processes with particle creation starting from a single particle. A
phase transition from a zero density absorbing state \cite{marro99}
can thus take place when $\tilde{\Gamma}_1$ changes sign. It is worth
noting that the transition threshold takes the same form as in
homogeneous MF theory, and it is thus independent of the network
topology, contrary to what is found in the bosonic SIS model
\cite{v.07:_react}, and similar to the case of the fermionic contact
process (CP) \cite{castellano06:_non_mean}. This is due to the fact
that SIS model is represented in terms of a two-species RD process,
see Eq.~(\ref{eq:25}), in which, moreover, a conservation rule (total
number of particles) is imposed. This conservation rule, coupled to
the diffusive nature of both species, is at the core of the zero
threshold observed in the SIS on SF networks in the
thermodynamic limit \cite{v.07:_react}. The contact process, on the
other hand, belongs (in Euclidean lattices) to the same universality
class as the one-species Schl\"{o}gl RD process
\cite{janssen81:_noneq_phase}, hence the topology-independent
threshold in the fermionic CP in networks can be understood in view of
the general result just derived in the bosonic framework.

To make further progress we restrict our attention to the case of
uncorrelated networks, in which \cite{dorogorev}
\begin{equation}
  P(k'|k) = \frac{k' P(k')}{\avk}.
\end{equation}
In this case, Eq.~(\ref{eq:4}) can be rewritten as
\begin{equation}
  \rho_k = - \frac{k \rho}{\avk \Gamma_1} - \sum_{q>1}
  \frac{\Gamma_q}{\Gamma_1}  [\rho_k]^q,  
\label{eq:23}
\end{equation}
Solving Eq.~(\ref{eq:23}), we find an expression $\rho_k(\rho)$,
depending implicitly on the particle density.  Inserting this solution
into Eq.~(\ref{eq:5}), we obtain a self-consistent equation for
$\rho$,
\begin{equation}
  \rho = \sum_k P(k) \rho_k(\rho),
  \label{eq:7}
\end{equation}
to be solved in order to obtain $\rho$ as a function of the RD
parameters.  

An approximate solution of Eq.~(\ref{eq:23}) can be obtained in the
limit of a very small particle density, that is, very close to the
threshold. In this case, we can neglect the higher order terms in
Eq.~(\ref{eq:23}) and obtain
\begin{equation}
  \rho_k \simeq - \frac{k}{\avk \Gamma_1} \rho,
  \label{eq:8}
\end{equation}
which makes only sense for $\Gamma_1<0$ (i.e. $0 < \tilde{\Gamma}_1
<1$, close to the phase transition). Inserting this expression into
the self-consistent equation (\ref{eq:7}) yields no information. We
must use, instead, the self-consistent relation coming from the
steady-state condition of Eq.~(\ref{eq:21}), namely
\begin{equation}
   \rho = -\frac{1}{\tilde{\Gamma}_1} \sum_{q>1} \Gamma_q \sum_k
   P(k)  [\rho_k]^q.\label{eq:24}
\end{equation}
Inserting (\ref{eq:8}) into Eq.~(\ref{eq:24}), and keeping only the
term corresponding to the reactions of lowest order $q_m>1$, we obtain
\begin{equation}
  \rho \simeq \left( \frac{(\avk |\Gamma_1|)^{q_m}}{ \langle k^{q_m}
      \rangle|\Gamma_{q_m}| } \right)^{1/(q_m-1)}  \;
  {\tilde{\Gamma}_1}^{1/(q_m-1)}, 
  \label{eq:26}
\end{equation}
where we have assumed $\Gamma_{q_m}<0$. This solution indicates that,
in a finite size network and for sufficiently small densities, all
bosonic RD systems with an absorbing state show a critical point
${\tilde{\Gamma}_1}^c =0$, with an associated density critical
exponent $\beta=1/(q_m-1)$, coinciding again with the homogeneous MF
solution. For SF networks with degree exponent $\gamma \leq
q_m+1$, the particle density is additionally suppressed by a diverging
factor $\langle k^{q_m}\rangle^{-1/(q_m-1)}$, signaling the presence
of very strong size effects. For $\gamma > q_m+1$, the particle
density is size independent, and we recover the standard MF solution
for homogeneous systems.

\subsection{Monotonously decaying  Bosonic RD processes}
\label{sec:mon-dec-proc}

As we have seen in Sec. \ref{sec:steady-state-rd-1}, a necessary
condition for a RD system to have a decaying density is to have
$\tilde{\Gamma}_1 <0$. In this case, since no steady states are
present, the full Eq.~(\ref{eq:20}) must be solved. One can proceed by
using a quasi-stationary approximation \cite{michelediffusion},
assuming $\partial_t \rho_k(t) \ll \rho_k(t)$, which will be correct
at low densities if $\rho_k(t)$ decays as a power law.  Thus,
neglecting the left-hand-side of Eq.~(\ref{eq:20}), we obtain again
Eq.~(\ref{eq:23}). Solving it and inserting the corresponding
expression of $\rho_k$ back into Eq.~(\ref{eq:21}), we have an
approximate equation for $\rho(t)$ that can give information about the
long time behavior of the RD process.

This procedure can be simplified when considering the limit of very
large time and very small particle density, where the concentration of
particles is so low that the RD process is driven essentially by
diffusion. In this diffusion-limited regime, it is possible to
estimate the behavior of the particle density, which turns out to be
independent of the correlation pattern of the network.  Let us
consider the limit case $\tilde{\Gamma}_1=0$, that is, in the absence
of one particle reactions. Then, in the limit $\rho_k \to 0$, linear
terms dominate in Eq.~(\ref{eq:20}) and we can write
\begin{equation}
  \frac{\partial  \rho_k(t)}{\partial t} \simeq - \rho_k(t) 
  + k \sum_{k'} \frac{P(k'|k)}{k'} \rho_{k'},
\end{equation}
that is, the density behaves as in a pure diffusion problem. The
situation is thus the following: the time scale for the diffusion of
the particles is much smaller than the time scale for two consecutive
reaction events, therefore at any time the partial density is well
approximated by a pure diffusion of particles
\cite{lovasz, noh04:_random_walks_compl_networ, rw_rings},
\begin{equation}
\rho_k(t)\simeq \frac{k \rho(t)}{\langle k \rangle},
\label{diffusion_limit}
\end{equation}
proportional to the degree $k$ and the total concentration of
particles, and independent of degree correlations. Inserting this
quasi-stationary approximation back into Eq.~(\ref{eq:21}), we obtain
\begin{equation}
   \frac{\partial  \rho(t)}{\partial t} \simeq \sum_{q>1}
   \frac{\Gamma_q \avn{q}}{\avk^q} \rho^q(t).
\end{equation}
For small $\rho$, this equation is dominated by the reactions of
smallest order $q_m$. Therefore, assuming $\Gamma_{q_m} < 0$, we
obtain the same decay in time as in the homogeneous MF theory,
\begin{equation}
   \rho(t) \sim \left(\frac{(q_m-1)|\Gamma_{q_m}| \langle
       k^{q_m}\rangle}{\avk^{q_m}}\right)^{-1/(q_m-1)} \; t^{-1/(q_m-1)},
   \label{eq:2}
\end{equation}
again depressed by a size factor $\langle k^{q_m}\rangle^{-1/(q_m-1)}$
for $\gamma<q_m+1$, and completely independent of the correlation
pattern.

If we were interested in the time behavior at intermediate densities, finally, 
the full Eq.~(\ref{eq:21}) with the quasi-stationary approximation must be
solved. This in general can only be done for uncorrelated networks.

\section{Applications}
\label{sec:applications}

In this Section we will apply the bosonic MF formalism developed above
to the study of two examples of one-species RD processes, the
branching-annihilating random walk and the diffusion-annihilation
processes, representative of the classes of steady-state and
monotonously decaying processes, respectively. For the sake of
comparison, we will review also the predictions of corresponding
fermionic MF theory, developed for an interacting particle system
defined to simulate the process under scrutiny.

\subsection{Steady-state processes: Branching-annihilating random
  walks}
\label{sec:steady-state-proc}

On of the simplest RD processes leading to a nontrivial steady state
is the generalized branching-annihilating random walk (BARW), defined
by the reactions \cite{odor04:_univer}
\begin{equation}
   \begin{array}{ccc}
     q A &\stackrel{\lambda}{\longrightarrow}& \emptyset\\
     A&\stackrel{\mu}{\longrightarrow}& (p+1) A 
  \end{array},\label{eq:12}
\end{equation}
that is, particles annihilate in $q$-tuples with a rate $\lambda$, and
produce a number $p$ of offspring with rate $\mu$. Homogeneous MF
theory predicts a continuous phase transition at $\mu_c=0$, with a
particle density in the active phase
\begin{equation}
  \rho \sim \mu^{1/(q-1)}.
  \label{eq:6}
\end{equation}
For the particular case $q=2$, the transition belongs to different
universality classes, according to the parity of the number of
offsprings $p$ \cite{odor04:_univer}. If $p$ is an odd number, it
belongs to the universality class of directed percolation
\cite{marro99}, the same as the CP. On the other hand, an even $p$,
for which the parity of the number of particles in conserved, leads to
a new, and different, universality class.

\subsubsection{Fermionic MF theory}
\label{sec:fermionic_steady}

When analyzing the process given by Eq.~(\ref{eq:12}), the limitations
of a fermionic approach become evident.  Indeed, reactions involving
more than two particles are difficult to describe in a fermionic
framework, even from a conceptual point of view.  In fermionic models
\cite{originalA+A,weber:046108,castellano06:_non_mean}, usually
diffusion and reactions are intimately linked, since particles jump
between nearest neighbors and interact upon landing on an occupied
vertex. Thus, when more than two particles are involved in a single
reaction, complex schemes have to be devised to represent the process,
schemes which, on the other hand, cannot be easily handled with
standard sequential algorithms.  Possible solutions could be the use
of auxiliary ``intermediate'' particles
\cite{chang07:_react_diffus_proces}, or the
design new algorithms that include the circumstance of different
particles diffusing at the same time, but it is easy to figure out
situations that would be potentially critical for such schemes
(e.g. what happens in a pure diffusive process when one particle tries
to move to an occupied vertex, while its starting point has been
occupied by other particle?).  On the other hand, it would be possible
to construct such reaction schemes by involving a particle and two or
more of its nearest neighbors, in a reaction step independent of
diffusion. Such formalism, although possible in principle, would be
nevertheless not general, since the number of reacting particles would
be limited by the connectivity of the considered vertex, and it would
also be more cumbersome to analyze from a MF perspective.

To allow for a consistent fermionic description, we will restrict our
attention to the particular case $q=2$, in which only binary
annihilation events are allowed, and that can be defined as a
fermionic interacting particle system given by the rules:
\begin{itemize}
\item Each vertex can be occupied by at most one particle
\item With probability $f$, a particle jumps to a randomly
  chosen nearest neighbor.
  \begin{itemize}
  \item If the neighbor is empty, the particle fills it, leaving the
    first vertex empty.
  \item If the neighbor is occupied, the two particles annihilate,
    leaving both vertices empty.
  \end{itemize}
\item With probability  $1-f$, the particle generates $p$
  offsprings. To do so:
  \begin{itemize}
  \item $p$ different neighbors are randomly chosen
  \item A new offspring is created on every selected vertex, provided this is empty
        (if it is already occupied, nothing happens).		
  \end{itemize}
\end{itemize}
In order to avoid problems with the offspring generation step, the
minimum degree of the network is taken to be $m \geq p$. We note that
this algorithm is not parity conserving, but we do not expect this to
be relevant in networks at MF level.

With this implementation of the fermionic BARW in complex networks, we
can see that the corresponding MF theory for the density spectrum
takes the form
\begin{eqnarray}
   \frac{\partial  \rho_k}{\partial t} &=& - f \rho_k - f k \rho_k
   \sum_{k'} \frac{1}{k'} P(k'|k) \rho_{k'} \nonumber \\
   &+& f k (1-\rho_k)
   \sum_{k'} \frac{1}{k'} P(k'|k) \rho_{k'} \nonumber \\ 
   &+& (1-f) k (1-\rho_k) \sum_{k'}  \frac{p}{k'} P(k'|k) \rho_{k'},
\end{eqnarray}
where $p/k'$ is the probability that one offspring of a particle in a
vertex of degree $k'$ arrives at a given nearest neighbor. For the
particular case of uncorrelated networks, this equation simplifies to
\begin{equation}
   \frac{\partial  \rho_k}{\partial t} = - \rho_k - \frac{k
     \rho}{\avk} \rho_k + (1-\rho_k) (1+\nu) \frac{k
     \rho}{\avk}, 
\end{equation}
where we have rescaled the time and defined $\nu = (1-f)p/f$. The
steady-state condition $\partial_t \rho_k=0$ yields the expression
\begin{equation}
  \rho_k = \frac{k (1+ \nu) \rho / \avk}{1 + k (2+
    \nu)  \rho/\avk}. 
\label{barw_ferm_densk}
\end{equation}
Application of the self-consistent condition $\rho=\sum_kP(k)\rho_k$
yields
\begin{equation}
  \rho = \sum_k  \frac{P(k) k (1+ \nu) \rho / \avk}{1 + k
    (2+ \nu) 
    \rho/\avk} \equiv \Psi(\rho).
  \label{eq:15}
\end{equation}
The condition for the existence of a nonzero solution, $\Psi'(0) \leq
1$, yields the threshold for the existence of a steady state 
\begin{equation}
  \nu > \nu_c =0 \;\Rightarrow \; f<f_c=1.
\end{equation}
In order to obtain the asymptotic behavior of $\rho$ as a function of
$\nu$ in infinite SF networks, we proceed to integrate
Eq.~(\ref{eq:15}) in the continuous degree approximation, replacing
sums by integrals and using the normalized degree distribution $P(k) =
m^{\gamma-1} (\gamma-1) k^{-\gamma}$, where $m$ is the minimum degree
in the network, to obtain
\begin{equation}
   \rho =
   \frac{1+ \nu}{2+ \nu} F[1, \gamma-1, \gamma,
   -\frac{\avk}{  m (2+ \nu) \rho}],
\end{equation}
where $F[a,b,c,z]$ is the Gauss hypergeometric
function~\cite{abramovitz}.  Expanding the hypergeometric function in
the limit of small $\rho$, close to the absorbing phase, we recover at
lowest order for $\gamma>3$ the homogeneous MF result $\rho \sim
\nu$. For $2<\gamma<3$, on the other hand, we obtain
\begin{equation}
  \rho \sim \nu^{1/(\gamma-2)},
  \label{eq:3}
\end{equation}
corresponding to an absorbing state transition, given by the control
parameter $\nu$, with zero threshold and a critical exponent
$\beta=1/(\gamma-2)$.

In any finite network this behavior is modified by finite size
effects. To analyze it, we define $\Theta=\sum_kkP(k)\rho_k/\avk$.
The equation for the total density becomes then
\begin{equation}
  \frac{\partial \rho}{\partial t}=\rho[\nu-(2+\nu)\Theta].
\end{equation}
By imposing stationarity ($\partial_t \rho =0$) and non-zero solution
($\rho\neq0$) one obtains
\begin{equation}
  \Theta=\frac{\nu}{(2+\nu)}.
\label{theta}
\end{equation}
The expression of $\rho_k$, Eq.~(\ref{barw_ferm_densk}), can be
simplified in the small density regime ($\rho \ll \avk/[k(2+\nu)]$,
$\forall k$) as
\begin{equation}
  \rho_k\simeq\frac{k(1+\nu)\rho}{\avk}.
\end{equation}
By substituting this expression in the definition of $\Theta$ and
inserting it into Eq.~(\ref{theta}) one obtains
\begin{equation}
  \rho=\frac{\avk^2}{\fluck}\frac{\nu}{(1+\nu)(2+\nu)}.
\end{equation}
SF networks of finite size have a cutoff or maximum degree
$k_c(N)$ which is a function of N\cite{dorogorev}. Therefore, for
uncorrelated SF networks with degree cutoff scaling with the
network size as $k_c(N) \sim N^{1/2}$, finite size effects in the
fermionic BARW lead to a size dependent density scaling as
\begin{equation}
  \rho \sim N^{\frac{-(3-\gamma)}{2}}\nu.
\label{barw_ferm_densstaz}
\end{equation}

\subsubsection{Bosonic  MF theory}
\label{sec:branch-annih-rand}

A bosonic formalism imposes no practical restriction to the maximum
order that the reaction steps may have. Thus, the general BARW
defined by the reactions Eq.~(\ref{eq:12}) yields, within the bosonic
MF formalism, to a rate equation Eq.~(\ref{eq:20}) with
$\tilde{\Gamma}_1= p \mu$ and $\Gamma_q= - q \lambda$, and
$\Gamma_{q'}=0$, for $q'\neq \{1, q\}$, corresponding to an absorbing
state phase transition at a critical particle creation rate
$\mu_c=0$. The full analysis of this equation for any $q$ can be
cumbersome, but we can immediately predict the behavior at large times
in finite networks, which will be given by Eq.~(\ref{eq:26}), namely
\begin{eqnarray}
  \rho &\simeq& \left( \frac{[\avk p (1-p \mu)]^{q}}{ \langle k^{q}
      \rangle q \lambda  } \right)^{1/(q-1)}  \;
  \mu^{1/(q-1)} \nonumber \\
   &\sim& N^{-\frac{q+1-\gamma}{2(q-1)}}  \; \mu^{1/(q-1)},
  \label{eq:30}
\end{eqnarray}
for uncorrelated networks. 

To proceed further, we consider the simplest case $q=2$, in which the
density spectrum fulfills the equation
\begin{equation}
  |\Gamma_2|  \rho_k^2 - \Gamma_1  \rho_k - \frac{k \rho}{\avk}=0,
\end{equation}
yielding the solution
\begin{equation}
  \rho_k = \frac{|\Gamma_1|}{2 |\Gamma_2| } \left(-1 + \sqrt{1+\frac{4
         |\Gamma_2| \rho k}{\avk |\Gamma_1|^2}} \right),
\label{eq:barw_spectrum}
\end{equation}
where, in order to ensure the existence of the absorbing state, we
must impose the condition $\Gamma_1<0$. 
In the large  $k\rho$ regime, we observe here a distinctively
square root dependence,
\begin{equation}
 \rho_k \simeq \sqrt{\frac{k \rho}{|\Gamma_2| \avk }},
\end{equation}
different from the limiting constant behavior observed in the
corresponding fermionic formulation, Eq.~(\ref{barw_ferm_densk}), as
well as in other fermionic models
\cite{pv01a,michelediffusion,castellano06:_non_mean}.  In the low
density regime, on the other hand, we can Taylor expand
Eq.~(\ref{eq:barw_spectrum}) and recover, for the particular case of
the BARW, the general relation Eq.~(\ref{eq:8}).  Thus, for particle
densities smaller than the crossover density $\rho_\times$, with
\begin{equation}
 \frac{4 |\Gamma_2| \rho_\times k_c}{\avk |\Gamma_1|^2} = 1,
\label{eq:27}
\end{equation}
we recover, for uncorrelated SF networks, the asymptotic
finite size solution for $q_m=2$, given by Eq.~(\ref{eq:30}).

For networks in the infinite size limit, introducing the density
spectrum of Eq.~(\ref{eq:barw_spectrum}) into the self-consistent
equation Eq.~(\ref{eq:7}), we obtain
\begin{equation}
  \rho=\sum_k P(k) \frac{|\Gamma_1|}{2 |\Gamma_2|} \left(-1 +
    \sqrt{1+\frac{4 
        |\Gamma_2| \rho k}{\avk (\Gamma_1)^2}} \right).
\end{equation}
In the continuous degree approximation, we have
\begin{eqnarray}
  \rho &=&  \frac{|\Gamma_1|}{2 |\Gamma_2| } \left( -1 +
    \frac{2(\gamma-1)}{2\gamma-3} 
    \sqrt{\frac{4  |\Gamma_2| m \rho }{\avk |\Gamma_1|^2}} \right. \times
  \nonumber \\ 
  &\times& \left. F[-\frac{1}{2},
\gamma   -\frac{3}{2},  \gamma-\frac{1}{2},
  -\frac{\avk |\Gamma_1|^2}{4 |\Gamma_2| m \rho}] \right).
\end{eqnarray}
Expanding $F[a,b,c,z]$ in the limit of small $\rho$, we
find 
\begin{eqnarray}
  \rho &\simeq& \frac{\rho}{|\Gamma_1|} +  \frac{|\Gamma_1|}{4
      |\Gamma_2| \sqrt{\pi}} \Gamma(2-\gamma)
     \Gamma(\gamma-3/2) \times \nonumber \\
     &\times&\left(\frac{4 m  |\Gamma_2|
         \rho}{\avk(\Gamma_1)^2 } \right)^{\gamma-1} +
     \mathcal{O}(\rho^2). 
\end{eqnarray}
At lowest order, and for $\gamma>3$, we recover the homogeneous MF
solution $\rho \sim \tilde{\Gamma}_1 \sim p \mu$. On the other hand,
for $2<\gamma<3$, the nonzero solution of this equation is
\begin{equation}
  \rho \sim \frac{\tilde{\Gamma}_1^{1/(\gamma-2)}}{|\Gamma_2|} \sim
  \frac{(p\mu)^{1/(\gamma-2)}}{\lambda}, 
  \label{eq:16}
\end{equation}
corresponding to an absorbing state transition, given by the control
parameter $\mu$, with zero threshold and a critical exponent
$\beta=1/(\gamma-2)$, in full agreement with the results for the
corresponding fermionic version of the model.  We can use this last
result to estimate the crossover density to the finite size solution
Eq.~(\ref{eq:30}). Inserting Eq.~(\ref{eq:16}) into Eq.~(\ref{eq:27}),
and considering $\Gamma_2$ as a constant, we obtain that the finite
size solution should be observed for a control parameter
\begin{equation}
  \mu < \mu_\times = \frac{k_c^{2-\gamma}}{p}.
  \label{eq:29}
\end{equation}
Therefore, for uncorrelated SF networks, finite size effects
in the bosonic BARW should appear for a particle creation rate smaller
that $\mu_\times \sim N^{-(\gamma-2)/2}$.

\subsection{Decay processes: Diffusion-annihilation process}
\label{sec:decay-proc-diff}

The simplest case in the class of monotonously decaying RD processes
corresponds to the general diffusion-annihilation process
\begin{equation}
  q A  \stackrel{\lambda}{\longrightarrow} \emptyset,
  \label{eq:1}
\end{equation}
which is the particular case of the BARW analyzed in
Sec.~\ref{sec:steady-state-proc} with $\mu=0$ (at the critical point).
The homogeneous MF solution predicts a decay of the particle density
\begin{equation}
  \rho(t)  \sim t^{-1/(q-1)}.
  \label{eq:31}
\end{equation}
In Euclidean lattices of dimension $d$, dynamical renormalization
group arguments \cite{theoryA+A} show that the behavior in
Eq.~(\ref{eq:31}) is correct for $d$ above the critical dimension
$d_c=2/(q-1)$. Below it, we have instead $\rho(t) \sim t^{-d/2}$, with
logarithmic corrections appearing at $d=d_c$.

\subsubsection{Fermionic MF Theory}
\label{sec:fermionic_decay}

As discussed in Sec.~\ref{sec:fermionic_steady}, in order to allow for
a consistent fermionic description, we will restrict our attention to
the binary diffusion-annihilation process with $q=2$, which can be
implemented as a fermionic interacting system obeying the following
rules \cite{originalA+A,michelediffusion}:
\begin{itemize}
\item Each vertex can be occupied by at most one particle
\item Each particle jumps with probability $f$ to a randomly
  chosen nearest neighbor.
\item If the neighbor is empty, the particle fills it, leaving the
  first vertex empty.
\item If the neighbor is occupied, the two particles annihilate,
  leaving both vertices empty.
\end{itemize}
This model was analyzed in detail in
Ref.~\cite{michelediffusion}. There it was observed that the rate
equation for the density spectrum reads, in uncorrelated complex
networks, 
\begin{equation}
   \frac{\partial  \rho_k}{\partial t} = -\rho_k + \frac{k}{\avk}
   (1-2 \rho_k) \rho,
\end{equation}
where the probability $f$ has been absorbed into a rescaling of time.
With a quasi-stationary approximation, the density spectrum at large
times takes the form 
\begin{equation}
  \rho_k(t) = \frac{k \rho(t) / \avk}{1 + 2 k \rho(t) / \avk},
\end{equation}
which yields as a final equation for the density of particles
\begin{equation}
  \frac{\partial  \rho}{\partial t} = -2 \frac{\rho^2(t)}{\avk^2}
  \sum_k P(k) \frac{k^2}{1 + 2 k \rho(t) / \avk}. 
  \label{eq:14}
\end{equation}
In finite networks, for times larger that $t>t_\times$, with $ k_c
\rho(t_\times) \simeq 1$, the denominator in Eq.~(\ref{eq:14}) can be
simplified to $1$, to obtain the limit behavior in finite size
networks
\begin{equation}
  \rho(t) \simeq \frac{\avk^2}{2 \fluck} \; t^{-1}.
\end{equation}
In a network of infinite size, the full Eq.~(\ref{eq:14}) must be
integrated. Within the continuous degree approximation, this equation
takes the form
\begin{equation}
   \frac{\partial  \rho}{\partial t} = -\rho(t) F[1, \gamma-2,
   \gamma-1, -\avk/2 m \rho(t)].
\end{equation}
Expanding the Gauss hypergeometric function for small $\rho$, we
obtain, for $\gamma>3$, the asymptotic long time behavior $\rho(t)
\sim t^{-1}$ while for $2 < \gamma <3$, one has
\begin{equation}
  \rho(t) \sim t^{-1/(\gamma-2)}.
\end{equation}

\subsubsection{Bosonic MF Theory}
\label{sec:diff-annih-proc}

The general diffusion-annihilation process defined by reaction
Eq.~(\ref{eq:1}) leads to the general rate equation Eq.~(\ref{eq:20}),
with parameters $\tilde{\Gamma_1}=0$, $\Gamma_q=- q \lambda$, and
$\Gamma_{q'}=0$, for $q'\neq \{1, q\}$. In finite networks and for large
times, the behavior of the particle density will be given by
Eq.~(\ref{eq:2}), i.e.
\begin{eqnarray}
   \rho(t) &\simeq& \left(\frac{(q-1)q \lambda  \langle
       k^{q}\rangle}{\avk^{q}}\right)^{-1/(q-1)} \; t^{-1/(q-1)},
   \nonumber \\
   &\sim& N^{-\frac{q+1-\gamma}{2(q-1)}}  \; t^{-1/(q-1)},
   \label{eq:11}
\end{eqnarray}
the last expression holding for uncorrelated SF networks.

Let us focus again in the simplest case $q=2$.  The rate equation for
the total density in uncorrelated networks takes the form
\begin{equation}
   \frac{\partial  \rho(t)}{\partial t} = - |\Gamma_2|  \sum_k
   P(k)\rho_k^2(t).
\label{eq:10}
\end{equation}
Applying the quasi-stationary approximation for the density spectrum, we
are led to the second order equation
\begin{equation}
  |\Gamma_2| \rho_k^2 + \rho_k -  \frac{k}{\avk} \rho =0,
\end{equation}
whose only positive solution is
\begin{equation}
  \rho_k = \frac{1}{2  |\Gamma_2|} \left( -1 + \sqrt{1 + \frac{4 |\Gamma_2|
        k}{\avk} \rho} \right).
  \label{eq:9}
\end{equation}
For a finite network with degree cut-off $k_c$, when the density is
smaller that 
\begin{equation}
  \rho_\times = \frac{\avk}{4 |\Gamma_2|} k_c^{-1}, 
\label{eq:28}
\end{equation}
we can Taylor expand Eq.~(\ref{eq:9}) to obtain the expression $\rho_k
\simeq k \rho / \avk$ and the asymptotic behavior given by
Eq.~(\ref{eq:11}).
On the other hand, for large $k$ and $\rho$, we obtain
\begin{equation}
  \rho_k \simeq \sqrt{\frac{ k \rho}{4  |\Gamma_2| \avk}},
\end{equation}
and we find again the peculiar square root behavior of the density
spectrum on $k$, distinctive from the fermionic prediction.

The general solution in the infinite network limit can be obtained in
this case by substituting the quasi-stationary approximation
(\ref{eq:9}) into Eq.~(\ref{eq:10}), to obtain
\begin{equation}
  \frac{\partial  \rho}{\partial t} = - \frac{1}{ |\Gamma_2|} (1+2
   |\Gamma_2| \rho) +  |\Gamma_2| \sum_k P(k) \sqrt{1 + \frac{4  |\Gamma_2|
        k}{\avk} \rho}.
\end{equation}
In the continuous degree approximation, and for SF networks,
we obtain in the infinite network size limit
\begin{eqnarray}
  \frac{\partial  \rho}{\partial t} &=& - \rho - \frac{1}{ 2 |\Gamma_2|} 
  + \frac{\gamma-1}{ |\Gamma_2|(2\gamma-3)}
  \sqrt{\frac{4  |\Gamma_2| m \rho }{\avk}} \times \nonumber\\
  &\times& F[-\frac{1}{2}, \gamma
  -\frac{3}{2},  \gamma-\frac{1}{2},
  -\frac{\avk}{4  |\Gamma_2| m \rho}].
\end{eqnarray}
Considering the limit of large times and small densities, we can
expand the hypergeometric function \cite{abramovitz}, to obtain
\begin{equation}
  \frac{\partial  \rho}{\partial t} \simeq  \frac{\Gamma(2-\gamma)
    \Gamma(\gamma-3/2)}{4  |\Gamma_2| \sqrt{\pi}} \left(\frac{4 m |\Gamma_2|
      \rho}{\avk} \right)^{\gamma-1} +  \mathcal{O}(\rho^2)
\end{equation}
for $2<\gamma<3$, whose solution is
\begin{equation}
  \rho(t) \sim |\Gamma_2|^{\gamma-2}t^{-1/(\gamma-2)} \sim
  \lambda^{\gamma-2}t^{-1/(\gamma-2)}
  \label{eq:13}
\end{equation}
that is, a power law decay with an exponent $1/(\gamma-2)$, again in
agreement with the fermionic implementation of the process.  From this
expression we can estimate the time at which the crossover density in
Eq.~(\ref{eq:28}) is reached in SF network, namely
\begin{equation}
  t_\times \sim k_c^{\gamma-2} \sim N^{(\gamma-2)/2},
\end{equation}
taking the same functional form as the crossover control parameter for
BARW in Eq.~(\ref{eq:29}).

\section{Numerical simulations}
\label{sec:numer-simul}

As we have seen in the previous Sections, heterogeneous MF theory
applied to steady state and monotonously decaying bosonic RD processes
can make general predictions for the asymptotic behavior at finite
networks, as well as give specific solutions for the infinite network
size limit. In particular, we have seen that bosonic formalisms
provide exactly the same results as their fermionic counterpart
(whenever the fermionic mapping is possible) regarding the evolution
of the particle density, the only difference being the form of the
density spectra as a function of the degree $k$. In order to check
these conclusions, we have performed extensive numerical simulations
of bosonic and fermionic versions of the processes considered. To generate the 
network substrate for the RD processes, we have adopted the
uncorrelated configuration model (UCM)~\cite{ucmmodel} that has the
double benefit of producing SF networks without degree
correlations \cite{mariancutofss,krzywickirandom} and with a tunable
degree exponent. When correlations were desired, the configuration
model (CM) \cite{bekessi72,benderoriginal,bollobas1980,molloy95} was
used, with the additional constraint of lack of multiple connections
and self-loops \cite{originnewman,maslovcorr,mariancutofss}.

Numerical simulations of fermionic RD processes must be tailored on a
case by case basis, depending on the specific interacting particle
system chosen to represent it
\cite{originalA+A,michelediffusion,weber:046108}.  As a general rule,
simulations are performed following a sequential Monte-Carlo
scheme~\cite{marro99}. At the beginning, $N\rho_0$ particles are
randomly distributed on the network, respecting the fermionic
constrain that at most one particle can be present on a single vertex,
i.e. $\rho_0 \leq 1$. Then, at time $t$, a particle is randomly
selected, and it undergoes the corresponding stochastic dynamics.  The
system is then updated according to the actions performed by the
selected particle, and finally time is increased as $t \to t+1/n(t)$,
where $n(t)$ is the number of particles at the beginning of the
simulation step. For bosonic processes, we have used a continuous time
formalism, details of which are given in the following subsection.

\subsection{Continuous time bosonic simulations}
\label{sec:cont-time-boson}

Previous approaches to the numerical simulation of bosonic RD
processes on complex networks~\cite{v.07:_react} relied on a parallel
updating rule in which reaction and diffusion steps alternate: after
all vertices have been updated for reaction, particles diffuse. This
approach, while feasible, must again be tailored in a case by case
basis, and strongly depends on the specific reactions of the process
under consideration.  Moreover, it introduces a subtle but relevant
problem as far as the density spectrum is concerned. Indeed, while
preserving the average density, pure diffusion immediately sets up the
characteristic linear behavior $\rho_k(t) \sim k$. Thus, the density
spectrum may assume (very) different aspects if we look at it after
the reaction step or after the diffusion one.  In order to overcome
these difficulties we have opted instead for a sequential algorithm,
which not only is absolutely general, but is in addition closer to the
spirit of the continuous time rate equations we have developed to
describe heterogeneous MF theory. The algorithm implemented is based
in the one proposed in Refs.~\cite{park,park2} for the case of regular
lattices. For one-species RD processes, the algorithm is described as
follows: In networks of size $N$, initial conditions for simulations
are usually a number $\rho_0 N$ of particles randomly distributed on
the network vertices, with no limitation on the occupation number of
single vertices.  To perform the dynamics, we consider the microscopic
configuration $\{\mathcal{C}\}$ of the bosonic system, which is
specified by the occupation number $n_i$ at each vertex $i$. A
standard master equation approach~\cite{vankampen} implies that, for
RD processes described by Eq.~(\ref{eq:RD_general}), the average
number of events in an infinitesimal time $dt$ is
\begin{equation}
E(dt,\{\mathcal{C}\}) = dt \sum_{i,r} \left( q^r ! \, \lambda_r + \frac{\delta(q^r,1)}{\sum_{r'}\delta(q^{r'},1)}  \right) \omega(n_i,q^r) 
\end{equation}
where 
\begin{eqnarray}
\omega(n_i,q^r) = \left(\begin{array}{c} 
n_i \\
q^r \end{array}
\right)
\end{eqnarray}
is the number of non-ordered $q^r$tuples of particles at vertex $i$.
Since the algorithm considers all reacting $q-$tuples as equivalent, it is convenient 
focusing on reaction orders $q$  rather than on specific reactions $r$. In general, 
a particular RD process defines 
a finite set $Q$ of allowed reaction orders $q$, that can be formally indicated as 
$Q = (\{ q\} | \, \exists r : q^r = q)$. 
At each time step one has to: (i) select a vertex $i$ (ii)
select the order $q$ of the candidate reaction (iii) determine which 
reaction $r$ occurs. In details:

\begin{description}
 \item[(i)] A vertex $i$ is selected with probability $W_i/M$, 
where $W_i=\sum_{q \in Q}  \omega(n_i,q)$ and $M=\sum_i W_i$; \\
\item[(ii)] A particular $q=q^*$ (with $q^* \in Q$) is selected with probability 
$ \omega(n_i,q^*) / W_i$; \\
\item[(iii)] A particular reaction $r$ of order $q^r = q^*$ occurs
with probability $q^r! \, \lambda^{r} \, \Delta t$,
where $\Delta t$ is a configuration-independent time constant. In case
$q^*=1$, in addition to reaction processes, the particle has the
diffusion option, which is chosen with probability $\Delta t$ 
(since we set the diffusion coefficient $D=1$).
\end{description}

\noindent Time is updated as $t \to t+ \Delta t / M$.
It is clear~\cite{park} that to have valid transition probabilities $\Delta t$ 
must be chosen so that the condition
\begin{equation}
\left( \delta(q,1) + q! \sum_{r:q^r=q} \,   \lambda_r
 \right) \Delta t 
\leq 1
\end{equation}
holds for all values of $q$. With this prescription an average
of $E(\Delta t,\{\mathcal{C}\})$ events occur in a time interval $\Delta t$.

\subsection{Branching-annihilating random walks}

\begin{figure}
  \centerline{
    \includegraphics*[width=0.45\textwidth]{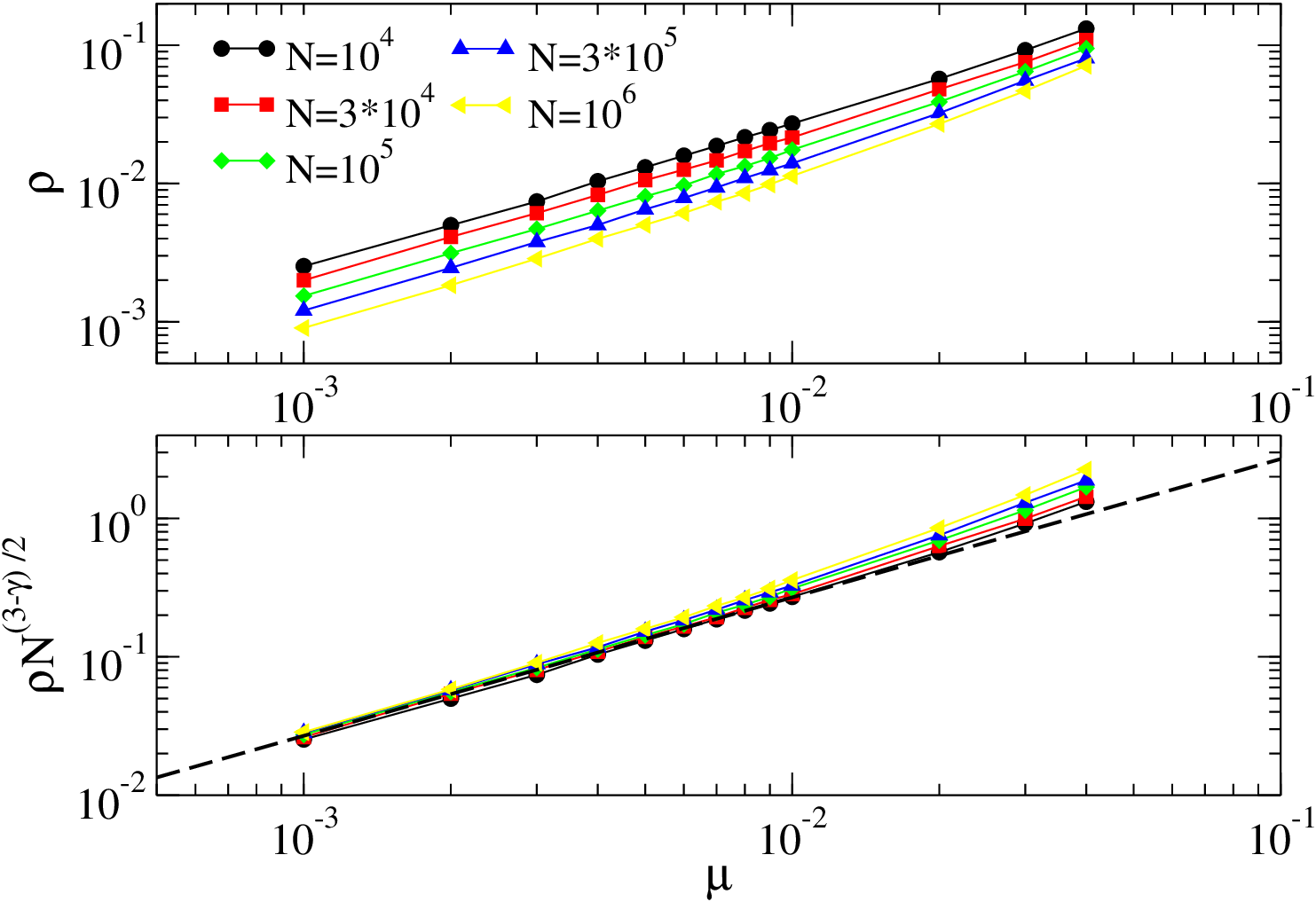}
  }
  \vspace*{0.5cm}
  \caption{Average density of the bosonic BARW with $q=p=2$ at the
    steady state on uncorrelated UCM networks with $\gamma=2.5$. The
    annihilation rate is kept fixed at $\lambda=0.1$.  Top: Density at
    the stationary state as a function of $\mu$, for different network
    sizes $N$. At any value of $\mu$, larger network sizes corresponds
    to smaller densities.  Bottom: Check of the collapse predicted by
    Eq.~(\ref{eq:30}). The dashed line has slope $1$.}
  \label{f:BARW_bos_density}
\end{figure}

In our numerical study of the BARW, we first focus in the behavior of
the average particle density in the steady state as a function of the
branching rate. As already observed in other dynamical systems in SF
networks \cite{michelediffusion,castellano07:_routes}, we find it
difficult to observe the infinite size limit behavior
[Eq.~(\ref{eq:3}) or~(\ref{eq:16})] in either bosonic of fermionic
simulations, for the network sizes available within our computer
resources. Therefore, we report the results for the finite size
behavior, expected in finite networks, Eqs.~(\ref{barw_ferm_densstaz})
and~(\ref{eq:30}). In Fig.~\ref{f:BARW_bos_density} we plot the
average density in the active phase of the bosonic BARW with $q=p=2$
as a function of the branching rate $\mu$.  In the parameter range
shown in this Figure (top panel), we observe that the density follows
a linear behavior as a function of the branching parameter $\mu$. This
linear dependence on $\mu$ corresponds to the asymptotic finite size
solution predicted by Eq.~(\ref{eq:30}), which is expected to hold in
networks of finite size and for very small steady state densities. We
can further check the accuracy of the prediction by noticing that, in
SF uncorrelated networks, the prefactor in $\rho$ should scale with
the system size as $\rho \sim \mu N^{-(3-\gamma)/2}$. Therefore, we
should expect that a plot of $N^{(3-\gamma)/2} \rho$ as a function of
$\mu$ would collapse for different network sizes. This is actually
what we observe in Fig.~\ref{f:BARW_bos_density} (bottom panel), where
different curves are clearly laid one on top of the other for small
values of $\mu$. As the density becomes larger, on the other hand, the
collapse becomes less and less precise, in agreement with the fact
that Eq.~(\ref{eq:30}) is only valid in the very small density regime.
Moreover, deviations from the collapse line set in earlier for large
system sizes in agreement with Eq.~(\ref{eq:29}), according to which
finite size effects show up for values of $\mu$ smaller than
$\mu_\times \sim N^{-(\gamma-2)/2}$.

\begin{figure}
\centerline{
\includegraphics[width=0.45\textwidth]{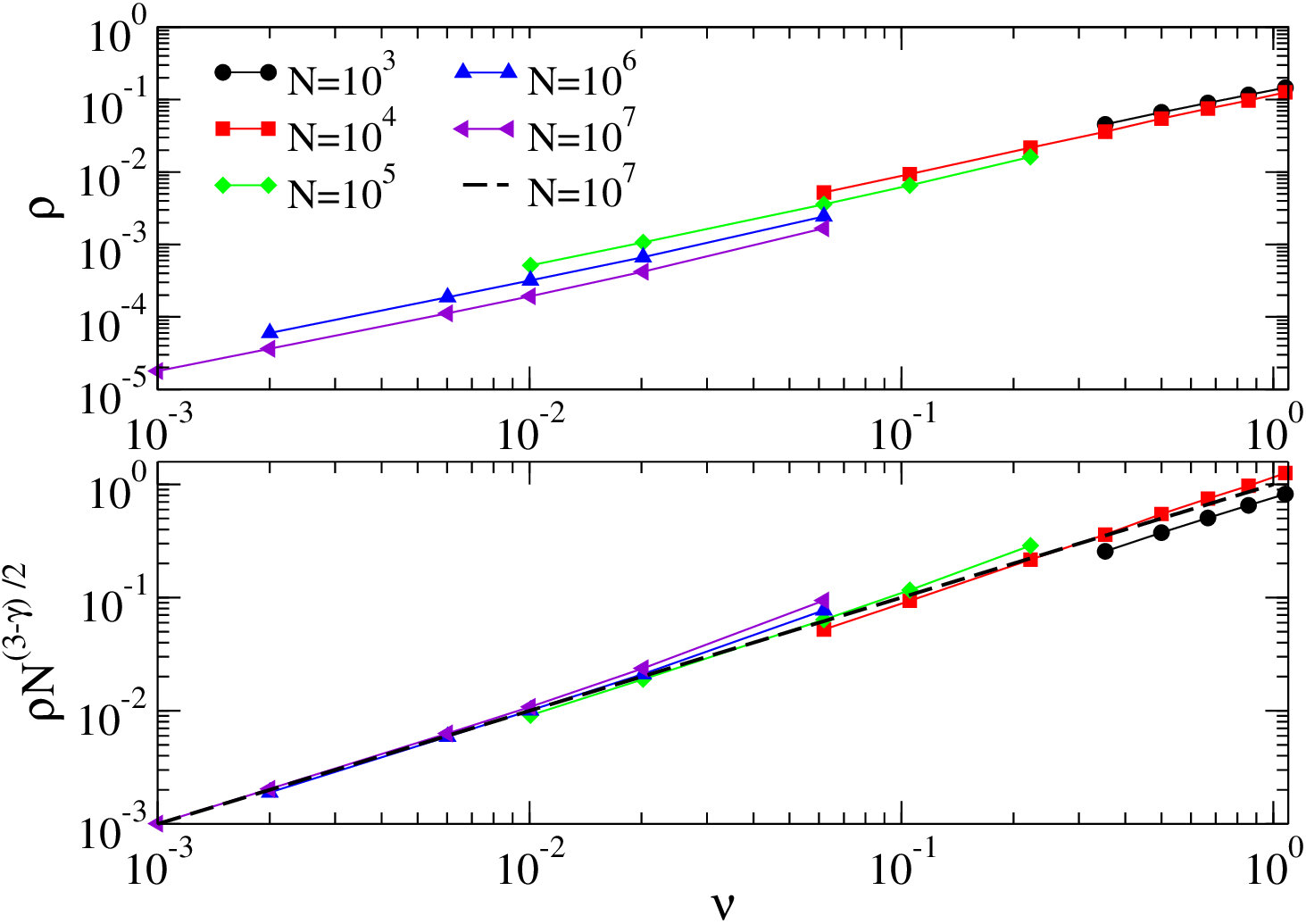} 
}
\vspace*{0.5cm}
\caption{Average density of the fermionic BARW with $q=p=2$ at the
  steady state on uncorrelated UCM networks with $\gamma=2.5$. Top:
  Density at the stationary state as a function of the parameter
  $\nu$, for different network sizes $N$. Bottom: Check of the
  collapse predicted by Eq.~(\ref{barw_ferm_densstaz}). The dashed
  line has slope $1$.}
\label{f:BARW_ferm_density}
\end{figure}

In Fig.~\ref{f:BARW_ferm_density} we present analogous results for the
fermionic version of the BARW. Here (top panel) we can observe a first
difference with respect to the bosonic BARW: For small values of $N$,
it is not possible to span a range of small values of $\nu$, due to
the fact the the system falls quickly into the absorbing state. Small
$\nu$ can only be explored using large $N$. The trend of all plots is,
however, correct: Linear in $\nu$ and decreasing when increasing the
network size.  The data again collapses with the same functional form,
now $\rho \sim \nu N^{-(3-\gamma)/2}$, for large systems sizes. The
deviations at small $N$ and large $\nu$, however, seem now larger than
in the bosonic case.

\begin{figure}
\centerline{
\includegraphics*[width=0.45\textwidth]{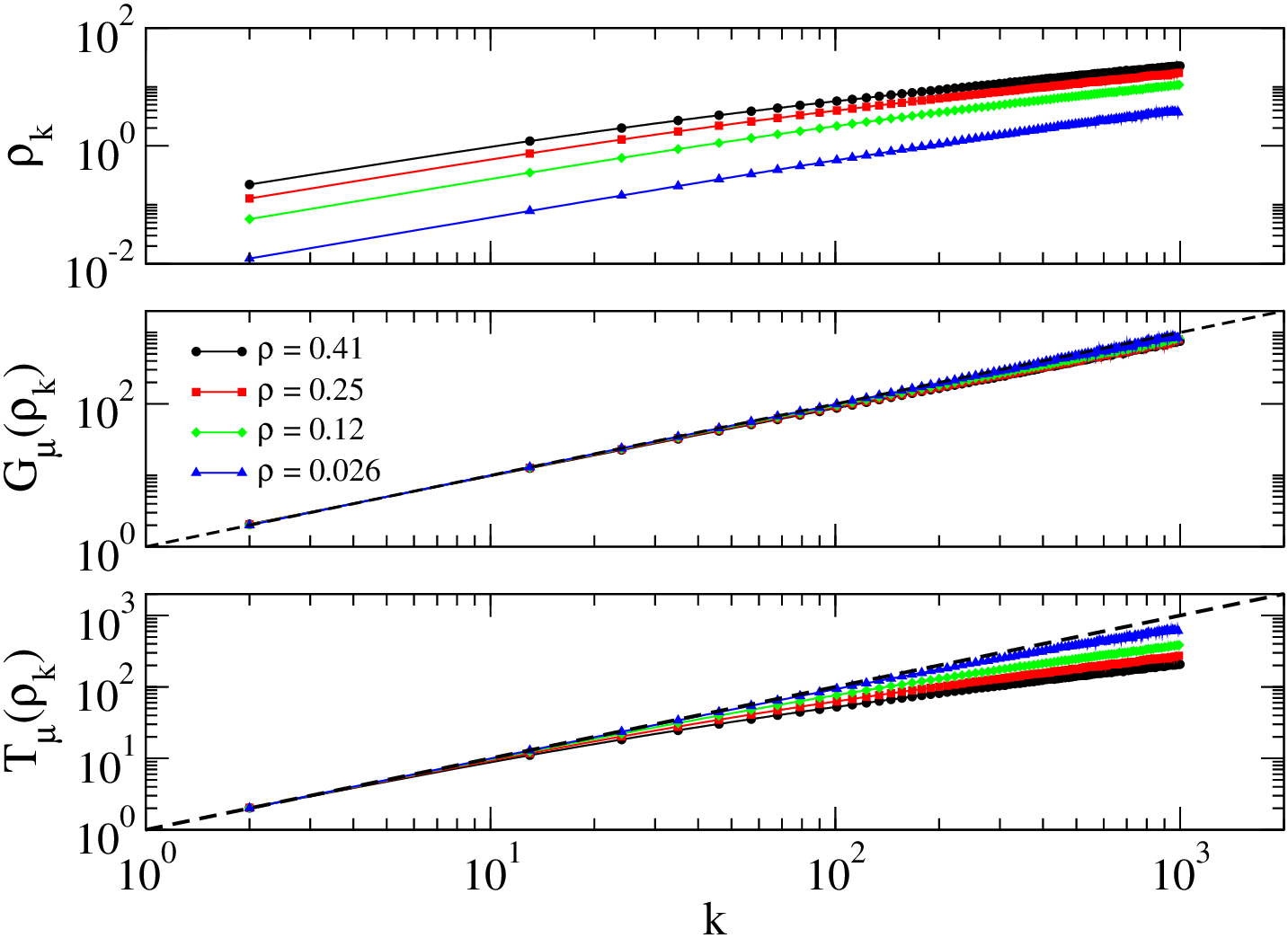} 
}
\vspace*{0.5cm}
\caption{Density spectra in the bosonic BARW process with $q=p=2$ at
  the steady state on uncorrelated UCM networks with
  $\gamma=2.5$. Network size $N=10^6$.  Top: Density spectra as a
  function of the degree $k$ for different steady state
  densities. Different stationary densities have been obtained fixing
  the annihilation parameter $\lambda=0.05$, and varying the branching
  parameter $\mu$.  Center: Data collapse of the density spectra with
  different average stationary densities  as predicted by
  Eq.~(\ref{eq:barw_spectrum_collapse}).  Bottom: Check of the Taylor
  expansion of the density spectra, as given by
  Eq.~(\ref{eq:barw_spectrum_taylor}).}
\label{f:BARW_bos_spectrum}
\end{figure}

Having checked that the average density takes the same form in both
bosonic and fermionic approaches, we focus now in the density spectra,
in which differences between the two formalisms are predicted at the
MF level. In the case of the bosonic BARW with $q=2$, the density
spectra in the steady state, as given by Eq.~(\ref{eq:barw_spectrum}),
is characterized by a peculiar square root behavior. To check this
form, we observe that, if we define the function
\begin{equation}
  G_\mu(\rho_k) \equiv \left[ \left( \frac{4\lambda \rho_k}{1-2 \mu} + 1 \right)^2
    -1 \right] \frac{(1-2\mu)^2 \langle k \rangle}{8 \lambda \rho},
  \label{eq:barw_spectrum_collapse}
\end{equation}
we expect $G_\mu(\rho_k) = k$ for any values of the reaction
parameters.  In Fig.~\ref{f:BARW_bos_spectrum}(center panel) we can
see that this collapse works well for a wide range of $\rho$
values. Alternatively, we can consider the small density behavior,
given by the general Eq.~(\ref{eq:8}), which translates in the
function
\begin{equation}
  \label{eq:barw_spectrum_taylor}
  T_\mu(\rho_k) \equiv (1-2\mu) \langle k \rangle  \frac{\rho_k}{\rho}
\end{equation}
being $T_\mu(\rho_k)=k$. In Fig.~\ref{f:BARW_bos_spectrum}(bottom
panel) we observe a poor collapse of the curves, which is
approximately attained only at very low densities, confirming the presence
of strong nonlinearities at large $\rho$.

\begin{figure}
\centerline{
\includegraphics*[width=0.45\textwidth]{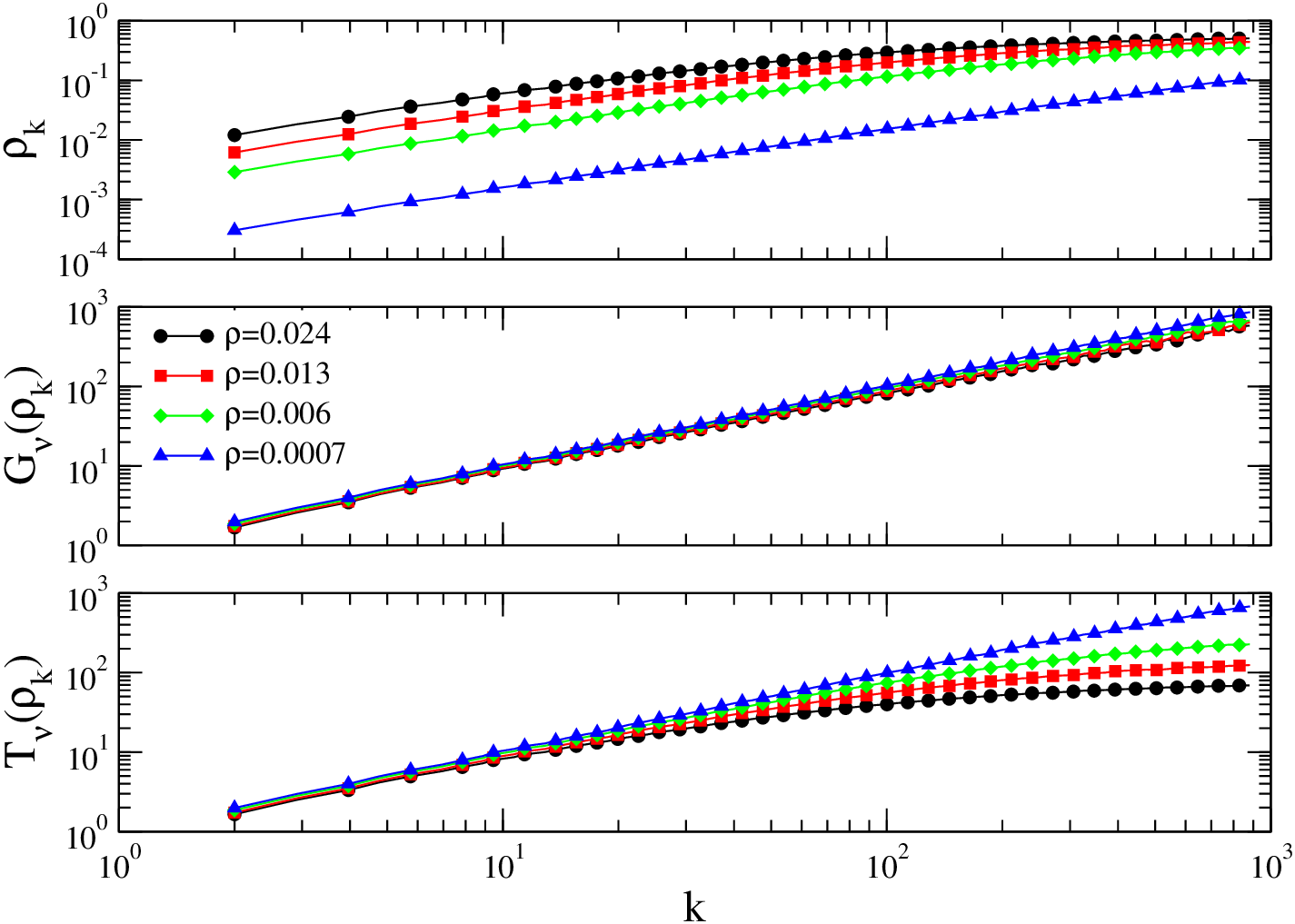} 
}
\vspace*{0.5cm}
\caption{Density spectra for the fermionic BARW process with $q=p=2$
  at the steady state on uncorrelated UCM networks with
  $\gamma=2.5$. Network size $N=10^6$.
  Top: Density spectra as a function of the degree $k$ for different
  steady state densities. Different stationary densities have been
  obtained varying the parameter $\nu$.  Center: Data collapse of the
  density spectra for different steady state densities, as predicted
  by Eq.~(\ref{eq:collapseferm}).  Bottom: Check of the Taylor
  expansion of the density spectra, as given by
  Eq.~(\ref{eq:taylorferm}).}
\label{f:BARW_ferm_spectrum}
\end{figure}

In Fig.~\ref{f:BARW_ferm_spectrum} we investigate the density spectrum
of a fermionic BARW for different values of the total density. As we
can observe (top panel), the spectra saturates to a constant value for
large values of $\rho$ and $k$, as expected from the theoretical
expression Eq.~(\ref{barw_ferm_densk}). On the other hand, this
equation implies that the function
\begin{equation}
  G_\nu(\rho_k) \equiv  \frac{\langle k \rangle \rho_k}{\rho(t) [(1+
    \nu) - (2+ \nu)\rho_k]} 
\label{eq:collapseferm}
\end{equation}
should satisfy $ G_\nu(\rho_k)=k$ for all $f$ and $p$.  Considering the
small density limit, on the other hand, a linear behavior of $\rho_k$
with $k$ is expected, translated again in the new function
\begin{equation}
  T_\nu(\rho_k) \equiv \frac{\langle k \rangle \rho_k}{\rho(t) (1+ \nu)}
\label{eq:taylorferm}
\end{equation}
being $ T_\nu(\rho_k)=k$.  While the collapse with the full shape of
Eq.~(\ref{barw_ferm_densk}) (center panels in
Fig.~\ref{f:BARW_ferm_spectrum}) is almost perfect, it is much worse
if only the Tailor expansion in considered (bottom panel), being only
approximately correct for very small densities.

\subsection{Diffusion-annihilation process}

\begin{figure}
\centerline{
\includegraphics*[width=0.45\textwidth]{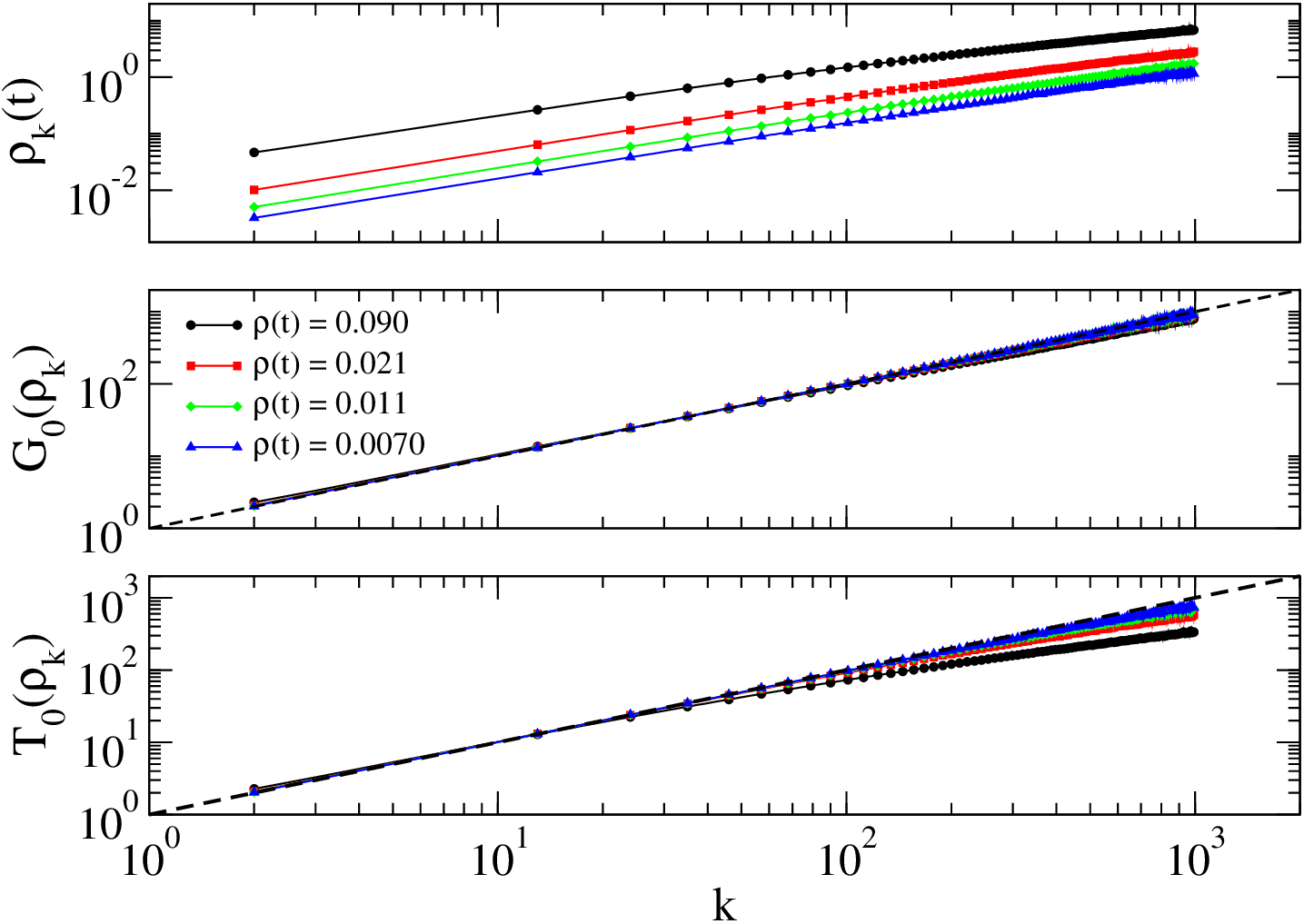} 
}
\vspace*{0.5cm}
\caption{Density spectra of the bosonic RD process $2A \to \emptyset$
  on uncorrelated UCM networks with $\gamma=2.5$.  Network size
  $N=10^6$.  Top: Density spectra as a function of the degree $k$ at
  different times (densities) from measures performed with fixed
  parameter $\lambda = 0.1$.  The curves show a bending in the large
  $k$ region for short times (large densities). Center: Data collapse
  of the density spectra at different times as predicted by
  Eq.~(\ref{eq:collapseAAbosonic}).  Bottom: Check of the Taylor
  expansion of the density spectra, as given by
  Eq.~(\ref{eq:taylorAAbosonic}). The poor collapse at large $k$
  confirms the presence of strong nonlinear terms at short times.}
\label{f:AA_bos_spectrum}
\end{figure}

To validate our theoretical approach for decaying RD systems, we have
concentrated on the bosonic description of the processes, since the
fermionic version described in Sec.~\ref{sec:fermionic_decay} was
already checked numerically in Ref.~\cite{michelediffusion}. We
consider thus the general bosonic process $q A \to \emptyset$, at rate
$\lambda$, for which a detailed analytical solution was given in
Sec. \ref{sec:diff-annih-proc}. For the case $q=2$, again a peculiar
square root behavior for the density spectrum was predicted in
Eq.~(\ref{eq:9}), which is corroborated in
Fig.~\ref{f:AA_bos_spectrum} by means of three different graphs.
Again, from Eq.~(\ref{eq:9}), defining the function
\begin{equation}
 G_0(\rho_k) \equiv \left( (4\lambda \rho_k  + 1) ^2  -1 \right) \frac{ \langle
k
\rangle}{8 \lambda \rho},
\label{eq:collapseAAbosonic}
\end{equation}
where $\lambda$ is the annihilation parameter, we will expect that
$G_0(\rho_k)=k$ for all times. In Fig.~\ref{f:AA_bos_spectrum} (center
panel) it is clear that the different curves, corresponding to
different values of the average density $\rho(t)$, collapse well in
agreement with the theoretical prediction.  In the bottom panel, we
check the general asymptotic expression for large times,
Eq.~(\ref{eq:8}). In this case, for small values of $8 \lambda k \rho
/ \langle k \rangle$, we should expect the function
\begin{equation}
 T_0(\rho_k) \equiv \frac{k \rho(t)}{\langle k \rangle}
\label{eq:taylorAAbosonic}
\end{equation}
to be $T_0(\rho_k)=k$, which holds when the times are large enough,
but shows a clear bending at large degrees and large densities,
signature again of the fact that it is fundamental to take into
account the non-linearity of the spectrum.

\begin{figure}
\centerline{
\includegraphics*[width=0.45\textwidth]{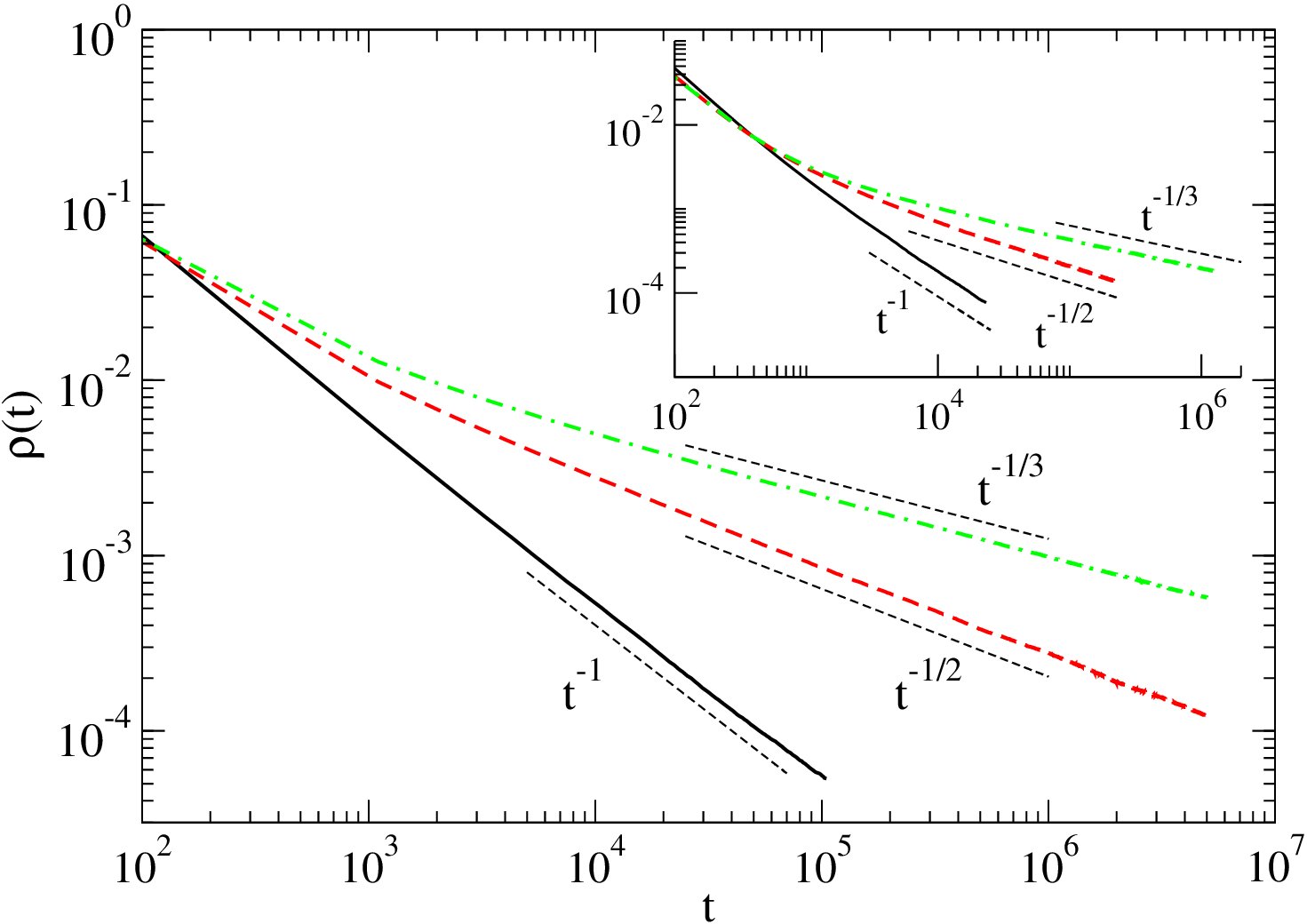} 
}
\vspace*{0.5cm}
\caption{Density decay of the bosonic $qA \rightarrow \emptyset$
  diffusion-annihilation processes in finite correlated and
  uncorrelated networks, for different values $q=2$ (full lines),
  $q=3$ (dashed lines) and $q=4$ (dot-dashed lines). For all values of
  $q$, the graphs show a tail of the form $t^{-1/(q-1)}$, as predicted
  in Eq.~(\ref{eq:11}), both for uncorrelated (UCM) networks (main
  figure) and correlated (CM) networks (inset). Data obtained from
  networks of size $N=10^5$ with degree exponent $\gamma=2.5$.  For
  all plots, the annihilation parameter was fixed at $\lambda=0.04$.}
\label{f:xA_0}
\end{figure}

As in the case of the fermionic diffusion-annihilation process
\cite{michelediffusion}, it turns out that the asymptotic expression
for infinite networks of the total particle density,
Eq.~(\ref{eq:13}), is very difficult to observe numerically, due to
the very small range of the extension of the power-law behavior.  We
have therefore focused again on the general prediction for finite
networks, Eq.~(\ref{eq:11}), according to which the RD process
$qA\to\emptyset$ shows a decay of the average density at large times
of the form $\rho(t) \sim t^{-1/(q-1)}$, independently of the presence
or absence of degree correlations.  We present in Fig.~\ref{f:xA_0}
simulation results for three values of $q$, namely $q=2,3,4$, on
uncorrelated networks SF generated with the UCM algorithm
(main plot), and correlated SF networks generated with the CM
prescription (inset). It is clear that the theoretical predictions are
in perfect agreement with numerical data.  This result is particularly
relevant since, for $q>2$, it concerns purely bosonic processes, which
do not have a fermionic counterpart.

\begin{figure}
\centerline{
\includegraphics*[width=0.45\textwidth]{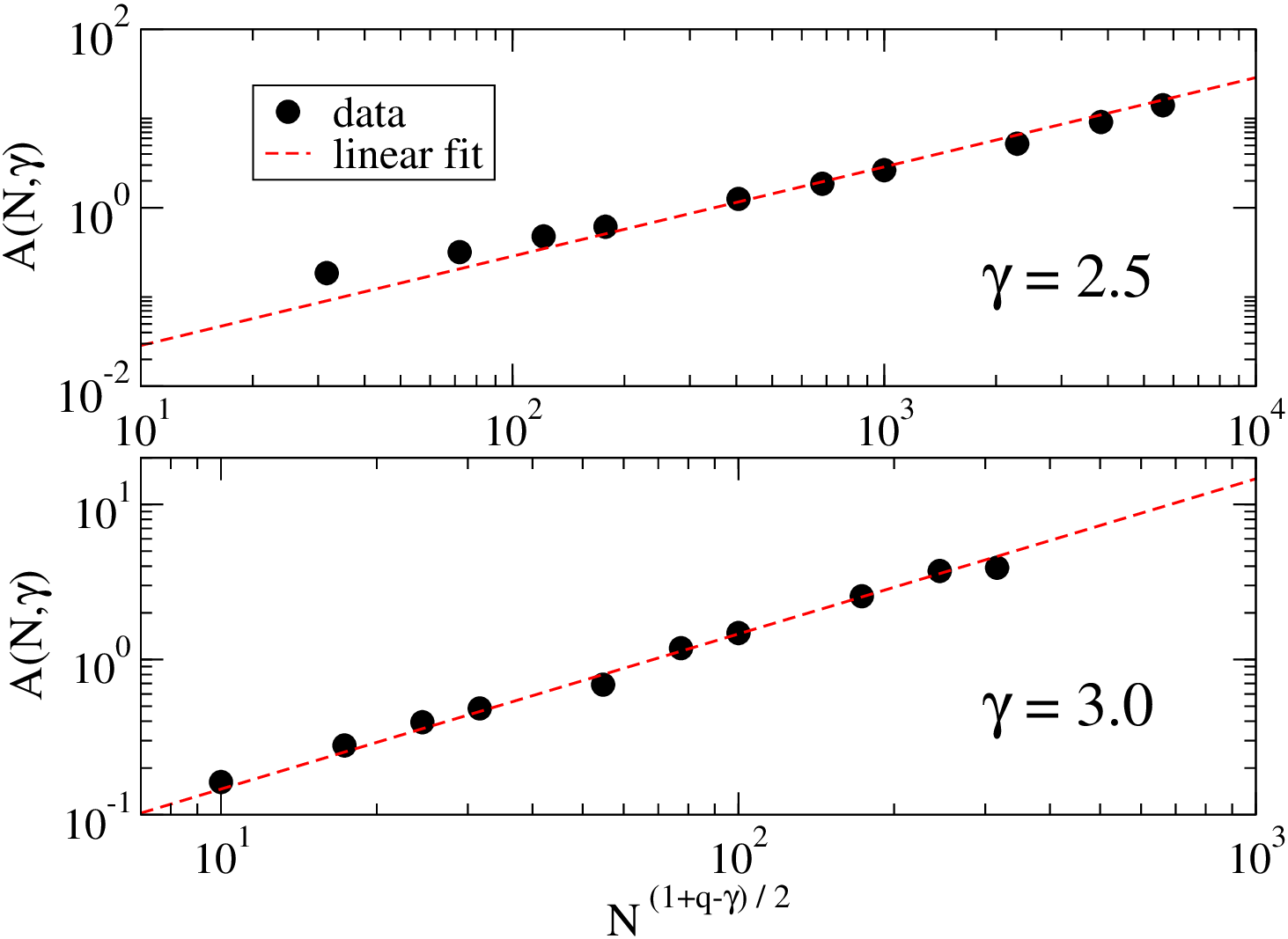} 
}
\vspace*{0.5cm}
\caption{System size dependence of the density prefactor in the
  diffusion-annihilation process $3A\to\emptyset$. According to
  Eq.~(\ref{eq:AA_scaling_prefactor}), we plot the prefactor
  $A(N,\gamma)$ as a function of $N^{(1 + q -
    \gamma)/2}$. The good linear behavior confirms the
  predictions of bosonic heterogeneous MF theory in the
  diffusion-limited regime.  Data obtained from networks with degree
  exponent $\gamma=2.5$.  For all plots, the annihilation parameter
  was fixed at $\lambda=0.04$.}
\label{f:AA_scaling_prefactor}
\end{figure}

The time independent prefactor of Eq.~(\ref{eq:11}), moreover, states
that the average density should be suppressed by the size term
$(\langle k^{q} \rangle / \langle k \rangle)^{-1/(q-1)}$. More
precisely, we can rewrite Eq.~(\ref{eq:31}) as $ \rho(t) \sim
A(N,\gamma)^{-1/(q-1)}t^{-1/(q-1)}$, with
\begin{equation}
A(N,\gamma) \sim N^{(1 + q - \gamma)/2}
\label{eq:AA_scaling_prefactor}
\end{equation}
in UCM networks, with cutoff $k_c(N) \sim N^{1/2}$.  We have estimated
the $A(N,\gamma)$ values by linear fits of the $\rho(t)^{-1}$ vs $t^{-1/(q-1)}$
curves for the $3A \rightarrow \emptyset$ process taking place on
networks of different sizes (data not shown), and for two values of
the degree exponent $\gamma$. We report the results in
Fig.~\ref{f:AA_scaling_prefactor}, where the scaling relation
predicted by Eq.~(\ref{eq:AA_scaling_prefactor}) is found to be in
very good agreement with simulation data.

\section{Discussion and conclusions}
\label{sec:discussion}

In this paper we have studied bosonic RD processes in SF networks
introducing a general continuous-time framework which is well suited
for both MF analytic calculations and computer simulations.  At the MF
level, we have developed the rate equations that characterize any
generic RD process.  We have considered in particular one-species RD
processes, for which MF theory provides a natural way to classify all
possible RD schemes. We have analyzed in detail both steady state and
monotonously decaying processes from a general perspective, focusing
also on specific examples, namely the BARW and 
diffusion-annihilation processes.  For processes characterized by
reactions not involving more than two particles, we have compared the
results with a fermionic version of the same problems, implemented in
terms of discrete interacting particle systems.

Beyond the obvious difference concerning the fact that the average
density is bounded in fermionic processes while it is not in their
bosonic version, both bosonic and fermionic MF formalisms render
equivalent results for the density of particles in single-species RD
processes, the main difference between both formalism lying in
functional form of the density spectrum of particles. For high
densities, the spectrum in bosonic systems goes in general as the
power $k^{1/q_M}$, where $q_M$ is the highest order of the reactions
defining the RD system, while for fermionic systems the behavior of
the spectrum is in general algebraic. Thus, in the bosonic scheme,
hubs become relatively less and less populated as more many-particle
reactions are present, provided the average density is sufficiently
high.  In the very low density regime, on the other hand, the bosonic
approach predicts spectra linear with $k$, a fact that allows to make
general predictions for the behavior of any RD process in finite
networks, which turns out to coincide with the homogeneous MF result,
with a network size correction. This results confirms the relevance of
finite size effects in dynamics on SF networks, already reported for
fermionic systems \cite{michelediffusion,castellano07:_routes}, since
the border between ``high'' and ``low'' densities is in general
determined by the $\rho k_c(N)$ product.

Another interesting result concerns one-species bosonic RD processes
with an absorbing state phase transition, where the critical point
does not depend on the possible heterogeneity of the network, but is
in general located at $\tilde{\Gamma}_1>0$. This condition translates
in the presence of reaction processes with particle creation starting
from a single particle and corresponds to the threshold independent of
the network topology found in other fermionic systems
\cite{castellano06:_non_mean}.  In order to observe effects of the
connectivity heterogeneity in the threshold, more complex RD schemes,
such as those involving two or more species, must be
considered~\cite{v.07:_react}. On the other hand, the bosonic point of
view allows to shed a different light on the value $\gamma=3$ usually
associated to a frontier between regular ($\gamma>3$) and complex
($\gamma<3$) behavior for dynamical systems on SF networks. We can
readily see that the value $\gamma=3$ emerges simply from considering
dynamical processes involving at most two particle interactions. For
general interactions involving $q$ particles, one will expect instead
to obtain unusual results for $\gamma<q+1$ (see e.g. Eq.~(\ref{eq:30})).

The continuous time theoretical and numerical formalisms presented for
bosonic processes have been developed in depth for the particular case
of one-species processes, but they can be easily generalized to many
species systems, opening thus the path to the study a large variety of
processes of large relevance in the understanding of the topological
effects of complex networks on dynamic and transport phenomena.

\section*{Acknowledgments}
We acknowledge financial support from the Spanish MEC (FEDER), under
projects No.  FIS2004-05923-C02-01 and No. FIS2007-66485-C02-01, and
additional support from the DURSI, Generalitat de Catalunya
(Spain). M.C. acknowledges financial support of Universitat
Polit\`ecnica de Catalunya.

\end{document}